\newcommand{\diracslash}[1]{#1\llap{/\kern2pt}}

\def\bearr{\begin{eqnarray}}
\def\eearr{\end{eqnarray}}

\newcommand{\be}{\begin{equation}}
\newcommand{\ee}{\end{equation}}
\newcommand{\bea}{\begin{eqnarray}}
\newcommand{\eea}{\end{eqnarray}}
\newcommand{\ba}[1]{\begin{array}{#1}}
\newcommand{\ea}{\end{array}}

\newcommand{\eqrf}[1]{Eq.\ (\ref{#1})}

\newcommand{\eqrftr}[3]{Eqs.\ (\ref{#1},\ref{#2},\ref{#3})}

\documentclass[preprint,secnumarabic,floats,amssymb,nobibnotes,nofootinbib,aps,prc,tightenlines]{revtex4}
\usepackage{graphicx}
\usepackage{amsmath}
\usepackage{latexsym}
\usepackage{epsfig}
\usepackage{subfigure}
\usepackage{multirow}
\usepackage{url}
\usepackage{float}
\usepackage{tabularx,booktabs}
\usepackage{xcolor}
\usepackage{hyperref}
\usepackage{array}
\newcolumntype{P}[1]{>{\centering\arraybackslash}p{#1}}

\begin{document}

\title{Non-canonical Higgs inflation }
\author{Pooja Pareek and Akhilesh Nautiyal}
\affiliation{Department of Physics, Malaviya National Institute of Technology Jaipur, 
JLN  Marg, Jaipur-302017, India}
\begin{abstract}
The large value of non-minimal coupling constant $\xi$ required to satisfy CMB observations
in Higgs inflation violates unitarity. 
In this work we study Higgs-inflation with non-canonical kinetic term of DBI form to find whether $\xi$ can be 
reduced. To study the inflationary dynamics, we transform the action to the Einstein frame, in which the Higgs is 
minimally coupled  to gravity with a non-canonical kinetic term and modified potential. We choose the Higgs self coupling constant
$\lambda=0.14$ for our analysis.
We find that the value of $\xi$ can be  reduced   from $10^{3}-10^{4}$ to $\mathcal{O}(10)$ to satisfy Planck constraints
on amplitude of scalar power spectrum. However, this model  produces a larger tensor-to-scalar ratio $r$, in comparison
to the Higgs inflation with canonical kinetic term. We also find that, to satisfy joint constraints on scalar spectral index  
$n_s$ and tensor-to-scalar ratio $r$ from Planck-2018 and bounds on $r$ from Planck and BICEP3, the value of $\xi$ should be
of the order of $10^4$. Thus, the  issue of unitarity violation remains even after considering Higgs inflation with
non-canonical kinetic term.
\end{abstract}


\maketitle

\section{Introduction}
Inflation \cite{Guth:1980zm} offers an explanation for the evolution of the universe and also addresses various 
cosmological issues with the hot Big Bang model, including the horizon problem and the flatness problem 
\cite{Linde:1983gd, Starobinsky:1980te}. During inflation the potential energy
of a  scalar field, named as inflaton,  dominates the energy density of the universe, which causes a quasi-exponential 
expansion.  At the time of inflation, the quantum fluctuations in the scalar field 
generate the primordial density perturbations, which leave  their  imprints in the large scale  structure (LSS) of the 
universe and temperature anisotropy in the cosmic microwave background (CMB) 
\cite{Mukhanov:1981xt,Starobinsky:1982ee,Hawking:1982cz}. There are also quantum fluctuations in the 
spacetime geometry during inflation generating primordial gravitational waves (tensor perturbations). 
The CMB and other LSS observations, specifically the most recent one from the Planck  satellite
\cite{Planck:2018vyg}, \cite{Planck:2018jri}, have placed significant  constraints on the various inflationary
 parameters. Despite the lack of a unique model of inflation, its predictions,  
such as nearly scale-invariant, Gaussian, and adiabatic density 
perturbations, are in excellent agreement with CMB observations.  The  power spectra of primordial density perturbations and 
tensor perturbations, generated during inflation, depend on the choice of inflaton potential. 
Any successful model of inflation should satisfy the two important criteria: (i) the scalar perturbations are ``well-behaved" 
during  inflationary phase, and (ii) there are natural methods to
terminate inflation ``gracefully". The  models of particle physics or string  theory 
can be used to determine the form of the inflaton potential \cite{Lyth:1998xn}. 


One of the best suited model of inflation from Planck-2018 observations \cite{Planck:2018jri}
is Higgs inflation \cite{Bezrukov:2007ep,Barvinsky:2008ia, Bezrukov:2010jz,Steinwachs:2019hdr, Rubio:2018ogq, Horn:2020wif}, where  the Higgs field of the standard model of particle physics is non-minimally coupled to gravity to achieve inflation.  
  The quartic potential for the minimally coupled Higgs field does not  fit well with  CMB observations,
however, the Higgs field coupled with gravity leads to a model that agrees very well with the current observations
 \cite{Bezrukov:2007ep,Bezrukov:2013fka}. The non-minimal coupling of the Higgs field and gravity is expressed as $\xi 
\phi^{\dag} \phi R $, where R is Ricci scalar and $\xi$ is non-minimal coupling constant. 
This non-minimal coupling term in the Lagrangian comes into existence due the quantum corrections 
to scalar field theory in curved spacetime, as it is necessary for the renormalization of the 
energy-momentum tensor \cite{Birrell:1982ix}. Non-minimal coupling of scalar field with gravity  can also
induce spontaneous symmetry breaking without having a negative sign of mass term \cite{Moniz:1990kt}. 
A dynamical system analysis for inflation with 
non-minimal coupling was performed in \cite{Barroso:1991aj}, and it was shown that inflation is possible for a
wide range of $\xi$. The advantages of considering non-minimal coupling is that, it 
helps the scalar field to exit smoothly at the end of inflation. 
If we consider standard model Higgs as inflaton with non-minimal coupling,  no  additional degree of freedom is required
between electroweak and Planck scale to be  consistent with CMB observations. 
To satisfy CMB observations, for $\lambda=0.14$, the value of the non-minimal coupling constant $\xi=10^4$.
For this large value of $\xi$ the non-minimal coupled Higgs inflation also faces  some theoretical 
problems. Due to this large value of $\xi$, the Higgs-Higgs scattering at a scale $\Lambda=M_{Pl}/\sqrt{\xi}$ via
graviton exchange becomes strongly coupled, which violates unitarity \cite{Burgess:2009ea, Barbon:2009ya, Burgess:2010zq, Lerner:2009na}. This does not affect the dynamics of inflation, but after inflation, 
at the time of preheating, when the Higgs inflaton starts to 
oscillate around the minima, a large number of longitudinal bosons are produced \cite{DeCross:2015uza, Ema:2016dny, 
Sfakianakis:2018lzf}. By adding extra scalar field or by introducing  $R^2$ term in Lagrangian, 
the unitarity of Higgs inflaton can be reestablished \cite{Ema:2017rqn, Giudice:2010ka, Lebedev:2011aq}. It is shown in \cite{Chakravarty:2013eqa} 
 that the non-minimal coupling constant $\xi \sim 1$ is allowed from Planck 
 observations for a generalization of Higgs inflation in $f(\phi, R)$ theory. 

Here we consider Higgs inflation with  non-canonical kinetic term to address the issue of unitarity violation.
 These models are named as  k-inflation 
\cite{Armendariz-Picon:1999hyi, Garriga:1999vw,Lambert:2002hk}, where inflation is achieved by the non-standard 
kinetic energy of inflaton rather than potential energy. 
The  non-canonical kinetic terms in the action of  inflaton field can be obtained from string theory
 \cite{Sen:1999xm, Sen:2000kd, Sen:2002nu}. 
The Dirac-Born-Infield form \cite{Gibbons:2002md,Alishahiha:2004eh, Ringeval:2009jd,Copeland:2010jt} or monomial and 
polynomial forms \cite{Mukhanov:2005bu} 
are the two possibilities for the non-canonical  kinetic terms in the action.
Various cosmological applications of  Born-Infeld theory with massive gauge fields have also been studied in 
\cite{VargasMoniz:2002ruz,Moniz:2002rd,VargasMoniz:2002gj}.
k-inflation introduces  new features in inflationary dynamics, such as a sound speed that is slower than the 
speed of light, which may also increase the non-gaussianity of the 
models \cite{Lidsey:2006ia}.  It also alters the predictions
for various inflationary parameters, such as scalar spectral index,
tensor-to-scalar ratio and running of the spectral index. k-inflation with the 
DBI form for kinetic energy along with quadratic, quartic and 
pseudo-Nambu-Goldstone boson (natural inflation) potentials is considered
in \cite{Devi:2011qm}, and constraints on various potential parameters are 
obtained from WMAP data. In \cite{Pareek:2021lxz} both the DBI form and 
the monomial and polynomial forms for the kinetic energy term along with 
polynomial potentials, PNGB potential and exponential potential have been 
analyzed in the light of reheating, and the constraints on various choices of
the kinetic term and potentials are obtained by demanding that the effective 
equation of state during reheating lies between $0$ and $1/3$ and the temperature
at the end of reheating is greater than $200$ GeV. Similar analysis is also 
done in \cite{Nautiyal:2018lyq} for tachyon inflation, where the potentials chosen are derived from 
string theory.
The non-canonical kinetic term, which causes inflation in the early universe, can also be used as the
 dark energy causing the late time acceleration in the Universe \cite{Chiba:1999ka,Armendariz-Picon:2000nqq, Armendariz-Picon:2000ulo, Chiba:2002mw,Chimento:2003ta, Chimento:2003zf}.
k-inflation with $f (R)$ gravity has been also considered in \cite{Nojiri:2019dqc}, 
In \cite{Odintsov:2021lum} k-inflation with inflaton coupled with a Gauss-Bonnet invariant is 
considered, and in \cite{Odintsov:2019ahz} k-inflation under constant-roll is studied. It has been shown in 
\cite{Gialamas:2019nly} that $R^2$ inflation in the Palatini  gravity  seems to be similar to k-inflation models in the 
Einstein frame.

The predictions for inflationary parameters for k-inflation  are consistent with the Planck-2018 measurements. 
In this work we consider Higgs potential with the DBI form for non-canonical kinetic term non-minimally coupled to gravity.
In our analysis we find the issue of unitarity violation still remains  for Higgs self-coupling
constant $\lambda=0.14$.
Higgs inflation with a monomial and polynomial form of non-canonical kinetic term was considered in \cite{Lee:2014spa};
 it was obtained that the inflaton remains sub-Planckian during inflation and hence, unitarity is not violated.  
The production of primordial black hole from Higgs inflation with a monomial and polynomial form of kinetic term was 
studied in \cite{Lin:2021vwc}.
Higgs inflation with non-minimal derivative coupling was  consider in \cite{Granda:2019wip}. 
Non-minimally coupled k-inflation with the DBI form for the kinetic term was considered in 
\cite{Piao:2002nh, Chingangbam:2004ng}, where the potential has an exponential and polynomial form derived from string 
theory. These models were further studied in the context of inflation as well as dark energy in \cite{Sen:2009fkx}. 

The paper is organized as  follows. We present a  general framework to analyze the behavior of a 
non-minimally coupled k-inflation in section \ref{nonminnoncan}. In section \ref{noncanhiggs}, we investigate 
the  Higgs inflation potential with non-canonical kinetic term and non-minimal coupling. We also find observational
constraints on various potential parameters from CMB and LSS observations in section \ref{noncanhiggs} Finally, Section 
\ref{conclusion} provides a summary of the findings from our analysis.

\section{Non-minimally coupled k-inflation : general framework} \label{nonminnoncan}
The action for non-minimally coupled  k-inflaton  is given as \cite{Piao:2002nh}, 
\cite{Sen:2009fkx}, \cite{Chingangbam:2004ng}  \\
\be
S = \int{d^4x}\sqrt{-g}\left\{\frac{M_{Pl}^2}{2}\left(1+\frac{\xi \phi^2}{M_{Pl}^2}\right)R -V\left(\phi\right)\sqrt{1+B g^{\mu\nu}\partial_{\mu}\phi\partial_{\nu}\phi}\right\}. 
\label{genaction}
\ee
Here, $R$ is the Ricci scalar, and $M_{Pl}$ represents the reduced Planck  mass, which is defined 
as $M_{Pl} = \frac{1}{8\pi G}$, where $G$ is Newton’s gravitational constant. 
In \eqrf{genaction}, $V(\phi)$ is the inflaton potential and $\xi$ is the gravity–field coupling constant. 
The action  (\ref{genaction}) illustrates non- minimally coupled non-canonical action in the Jordan frame.
In this action, field  $\phi$ has a dimension of mass, and the parameter $B$, which is introduced to make equation 
dimensionally consistent, has a dimension of $(mass)^{-4}$.

The action (\ref{genaction}) can be converted from the Jordan frame to the Einstein frame by 
eliminating the non-minimal coupling term, which requires a conformal transformation.
\be
\tilde{g}_{\mu\nu}\rightarrow F\left(\phi\right)g_{\mu\nu},
\label{conftrans}
\ee
where $F\left(\phi\right)= 1+\frac{\xi \phi^2}{M_{Pl}^2}$. The transformed action has the form

\be
S = \int{d^4x}\sqrt{-\tilde{g}}\left\{\frac{M_{Pl}^2}{2}\left(\tilde{R}-\frac32 \frac{F'^2}{F^2} \tilde{g}^{\mu\nu}\partial_{\mu}
\phi\partial_{\nu}\phi\right) -\tilde{V}\left(\phi\right)\sqrt{1+B F \tilde{g}^{\mu\nu}\partial_{\mu}\phi\partial_{\nu}\phi}\right\}. 
\label{transaction}
\ee
In this \eqrf{transaction}, $\tilde{R}$ represents the Ricci scalar in the Einstein frame. $ F'= \frac{dF(\phi)}{d\phi}$ is the first derivative of $F(\phi)$ with respect to field $\phi$, and 
$\tilde{V}$ is the  effective potential in the Einstein frame given as
\be
\tilde{V}\left(\phi\right)=\frac{V\left(\phi\right)}{F\left(\phi\right)^2}. \label{vtildeens}
\ee 
The action (\ref{transaction}) is comparable to the action for the minimally coupled k-inflation 
 with effective  potential  $\tilde{V}\left(\phi\right)$ except the term $\frac32 \frac{F'^2}{F^2}\tilde{g}^
{\mu\nu}\partial_{\mu}\phi\partial_{\nu}\phi$, which emerges as a result of the non-minimal coupling.

For the action with non-minimal coupling and having a canonical kinetic term,  the action in the Einstein frame is 
similar to the action in the minimally coupled scalar field, except that the shape of the potential changes due to 
the field redefinition. However, with the non-canonical kinetic term in the Jordan frame, 
 an additional term appears in the Einstein frame.  
This form of non-minimal coupling, in particular, always causes a correction to the 
Lagrangian ${\cal L}(X,\phi)$, where $X$ represents the   kinetic term  $X = \frac12 \partial_{\mu}\phi \partial^{\mu}\phi$.
It is important to note   that these models are stable  under quantum corrections as long as the conformal factor
 $F(\phi)$ and its variation are significantly  small.

In Einstein frame, it is simple to compute the energy density, pressure, and the equation of motion. 
The Friedmann-Robertson-Walker (FRW) metric is considered, which  has  the signature $(-,+,+,+)$, to evaluate the energy density $\rho$ and pressure $p$ for effective action (\ref{transaction}). The expressions for the energy density $\rho$ and pressure $p$ are obtained as
\bea
\rho &=& \frac{\tilde{V}}{\sqrt{1-B F \dot{\phi}^2}} + \frac34 M_{Pl}^2 \frac{F'^2}{F^2}{\dot{\phi}}^2, \label{rhotrans}\\
p &=& -\tilde{V}\sqrt{1-B F \dot{\phi}^2} + \frac34 M_{Pl}^2 \frac{F'^2}{F^2}{\dot{\phi}}^2. \label{ptrans}
\eea
The equation of motion for effective action (\ref{transaction}) can be obtained by varying it with respect to $\phi$ as
\be
\begin{split}
\left\{\frac{1}{1-B F \dot{\phi}^2} + \frac32 M_{Pl}^2 \left(\frac{F'}{F}\right)^2  \frac{\sqrt{1-B F
\dot{\phi}^2}}{B F \tilde{V}}\right\}\ddot{\phi} + \left\{1+\frac32 \left(\frac{F'}{F}\right)^2 M_{Pl}^2 \frac{\sqrt{1-BF
\dot{\phi}^2}}{B F \tilde{V}}\right\}3H\dot{\phi} \\
+\frac12 \left\{\frac{1}{1-B F \dot{\phi}^2}+3M_{Pl}^2\left(\frac{F''F-F'^2}{F^2}\right)\frac{\sqrt{1-B F \dot{\phi}^2}}{B
F \tilde{V}}\right\}\left(\frac{F'}{F}\right) \dot{\phi}^2+\frac{\tilde{V'}}{B F \tilde{V}} = 0. 
\end{split}
\label{eqmtrans}
\ee
The Friedmann equations, which describe the evolution of the universe, are 
\bea
H^2 &=& \frac{1}{3M_P^2}\rho, \label{hsquare}\\
\dot H &=& - \frac{1}{2M_P^2} \left(\rho+p\right). 
\label{hdot}
\eea

These equations for energy density  (\ref{rhotrans}) and pressure (\ref{ptrans}) can be expressed as
\bea
H^2  &=& \frac{\tilde{V}}{3\sqrt{1-B F \dot{\phi}^2}} + \frac14 M_{Pl}^2 \frac{F'^2}{F^2}{\dot{\phi}}^2, 
\label{hsquaref}\\
\dot H &=& -\frac{1}{2M_{Pl}^2}\left\{\frac{\tilde{V} B F\dot{\phi}^2}{\sqrt{1-B F \dot{\phi}^2}} + 
\frac32 M_{Pl}^2 \frac{F'^2}{F^2}{\dot{\phi}}^2\right\}. \label{hdotf}
\eea
For $F\left(\phi\right)=1$, \eqrf{hsquaref}, \eqrf{hdotf}, and \eqrf{eqmtrans},  reduce to the corresponding equations
for minimally coupled k-inflation. For $F\left(\phi\right)\neq 1$  potential is redefined to be $\tilde{V}$. 
The equation of motion, (\ref{eqmtrans}) contains  numerous terms, proportional to 
$\left(\frac{F'}{F}\right)^2$ and $\dot{\phi}^2$. Considering $\delta= \frac{M_{Pl}^2}
{B F \tilde{V}}\left(\frac{F'}{F}\right)^2\ll 1$, these terms can be ignored. With this the final expressions 
 for the equation of state and the Friedmann equations becomes
\bea
3H\dot\phi&=&-\frac{\tilde{V}'}{B F \tilde{V}},
\label{eqmaprox}\\
H^2 &=& \frac{\tilde{V}}{3M_{Pl}^2},\\
\label{hsquareaprox}
\dot H &=& -\frac{\tilde{V} B F\dot{\phi}^2}{2M_{Pl}^2}.
\label{hdotaprox}
\eea
The first two Hubble slow roll parameters $\epsilon_1$ and $\epsilon_2$  describing the dynamics of inflation 
can be expressed in terms of energy density and pressure as
\bea
&&\epsilon_1=\epsilon = \frac{3}{2}\frac{\rho+p}{\rho}\label{ep1gen}\\
\text{and}\;\; &&\epsilon_2 = \frac{3}{2H}\frac{d}{dt}\left(\frac{\rho+p}{\rho}\right)
\label{ep2gen}.
\eea
For energy density (\ref{rhotrans}) and pressure (\ref{ptrans}) the expressions for these two slow-roll parameters can be 
obtained as  
\bea
\epsilon_1 &=& \frac32 B F \dot{\phi}^2 = \frac{M_{Pl}^2} {2B} \frac{\tilde{V}'^2}{\tilde{V}^3},
\label{ep1}\\
\epsilon_2 &=& \frac{2\ddot{\phi}}{H \dot{\phi}}= \frac{M_{Pl}^2}{B F}
\left(\frac{F' \tilde{V}'}{F\tilde{V}^2}+3\frac{\tilde{V'}^2}{\tilde{V}^3}-2\frac{\tilde{V}''}{\tilde{V}^2}\right).
\label{ep2}
\eea
Here $\prime$ denotes the derivative with respect to field $\phi$.

According to slow roll conditions $|\epsilon_1|\ll 1$ and $\epsilon_2\ll 1$. The value of the field 
at the end of inflation can be  obtained by putting $|\epsilon_1(\phi_{end})=1|$.\\ 
The number of e-foldings 
during inflation is given by:
\be
N(\phi) = \int H\,dt \simeq \frac{B F}{M_{Pl}^2}\int_{\phi_e}^{\phi}\frac{\tilde{V}^2}{\tilde{V}'}\,d{\phi} \label{efoldphi}
\ee
Using this expression one can compute the number of e-foldings $N_k$ from the end of inflation to time when the length scales
corresponding to the pivot scale $k_0$ leave the Hubble radius during inflation.

 For our analysis we also require the Friedmann equation  (\ref{hdotaprox}), and equation of 
motion (\ref{eqmaprox}) in terms of number of e-foldings $N= \ln a$ as an independent variables. 
The equations obtained are
\be
\frac{dH}{dN}= \frac{H B V F}{2}\left(\frac{d\phi}{dN}\right)^2 
\label{hprime}
\ee
and
\be
H^2\frac{d^2\phi}{dN^2}+\left\{H \frac{dH}{dN}+3H^2\right\}\frac{d\phi}{dN}+\frac{1}{B F V}\frac{dV}{d\phi}=0. \label{eqmn}
\ee

The scalar and tensor power spectra in terms of Hubble parameters and potential, are given as \cite{Garriga:1999vw}
\bea
P_{S}(k) &=& \frac{H^2}{8\pi^2M_{Pl}^2c_s\epsilon_1} = \frac{1}{24 \pi^2 M_{Pl}^4 c_s}\frac{\tilde{V}}{\epsilon_1}
\label{scapwr}\\
P_h(k)&=&\frac{2H^2}{\pi^2 M_{Pl}^2} = \frac{2\tilde{V}}{3\pi^2 M_{Pl}^4}
\eea
In the above equations, perturbations are evaluated at the horizon crossing $c_s k=a H$. 
The value of scalar power spectrum for $k=k_0$, where $k_0=0.05$Mpc\textsuperscript{-1} is the pivot scale, 
provides the amplitude of scalar perturbations $A_S$. The  spectral index $n_s$, tensor-to-scalar ratio $r$ and the 
tensor spectral index $n_T$ are given  in terms of  
slow roll parameters as follows:
\bea
n_s-1 &=& \frac{d lnP_{S}(k)}{dlnk}= -2\epsilon_1-\epsilon_2 \label{nsv}\\
r &\equiv& \frac{P_{h}(k)}{P_S(k)} = 16 c_s \epsilon_1 \label{rv}\\
n_T &=& \frac{d lnP_h(k)}{dlnk}= -2\epsilon_1 \label{ntv}
\eea
where
\be
 c_S^2=\frac{\partial p/\partial X}{\partial \rho/\partial X} \label{cs2} 
\ee

is the sound speed. The value of $n_s$ and $r$ is also evaluated at the pivot scale $k_0$. In this work, we consider the Higgs inflation potential with a non-canonical kinetic term. 
for which the effective  potential will be $\tilde{V} \left(\phi\right)=\frac{V\left(\phi\right)}{F\left(\phi\right)^2}$ 
in the Einstein frame. The model and constraints are described in the next section. 

\section{Non-canonical Higgs inflation } \label{noncanhiggs}
\begin{figure}[H]
\includegraphics[height=6cm,width=7.2cm]{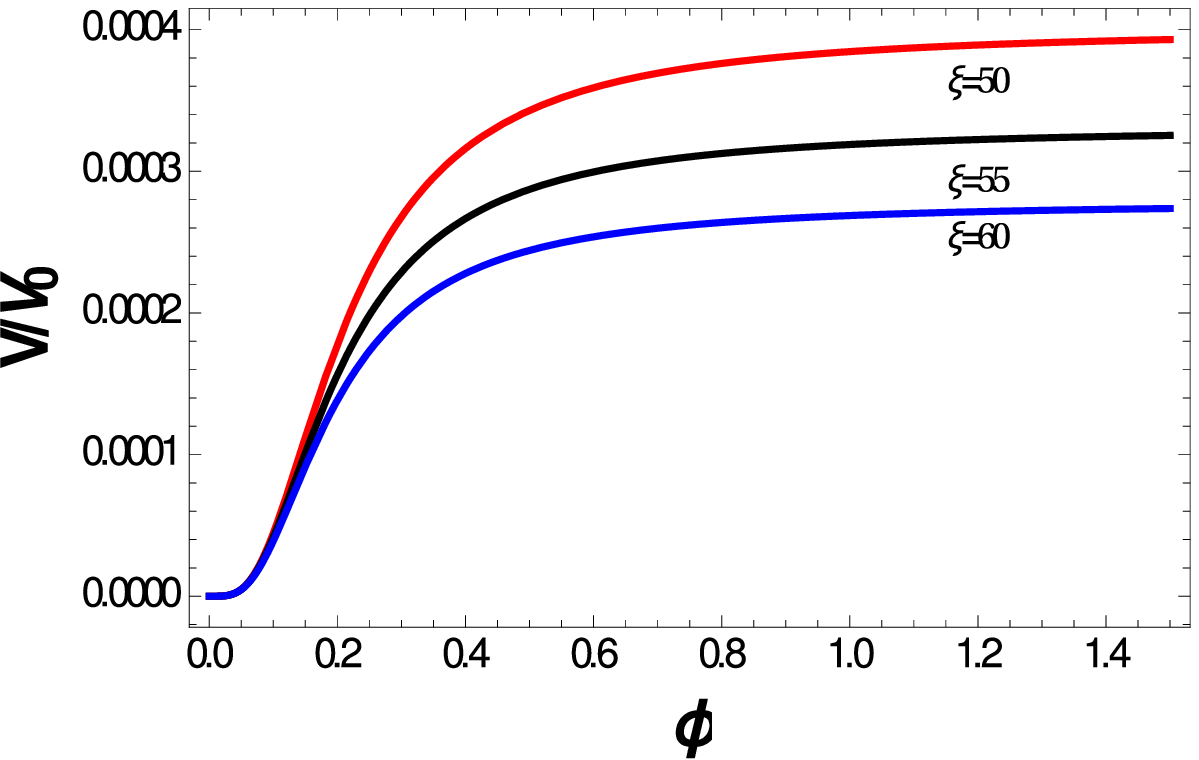}
\hspace{1cm}
\includegraphics[height=6cm,width=7.2cm]{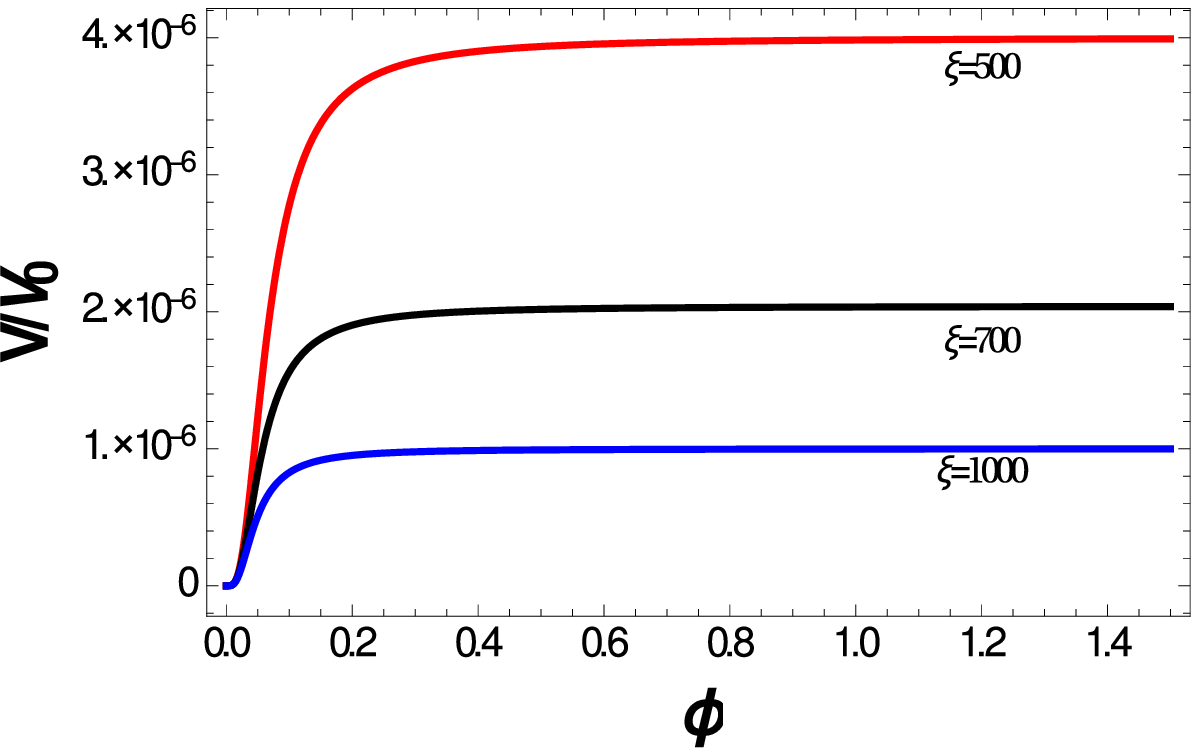}
\caption{In both panels potential $(V(\phi)/V_0)$ is plotted against field $\phi$, where $V_0 = \frac{\lambda}{4}$. 
The left panel corresponds to small values of $\xi$, where red, black and blue lines represents $\xi= 50,55$ and $60$. However, 
the right panel corresponds to large value of $\xi$, where red, black and blue lines represents  $\xi= 500,700$ and $1000$.}
\label{fig:potphi}
\end{figure}

The Higgs  potential  has the form \cite{Bezrukov:2007ep}
\be
V(\phi) = \frac{\lambda}{4}(\phi^2-\nu_{EW}^2)^2 \label{pothiggs}
\ee 
Here $\lambda$ is the self-coupling constant and $\nu_{EW}$ is the Higgs vacuum expectation value. 
The Higgs vacuum expectation value at  the  electroweak scale $\nu_{EW}= 246 GeV$ and $\nu_{EW}\ll\phi$.\\
In the non-minimally coupled non-canonical model, the potential is redefined in the Einstein frame. 
The expression of the redefined potential (\ref{vtildeens}) for Higgs potential (\ref{pothiggs}) becomes
\be
\tilde{V}(\phi) = \frac{\lambda}{4}\frac{\phi^4}{F(\phi)^2} =\frac{\lambda}{4}\frac{\phi^4}{\left(1+\frac{\xi \phi^2}{M_{Pl}^2}\right)^2}.
\label{redpot}
\ee
The redefined potential \eqrf{redpot}, exhibits the behavior shown in Fig.~\ref{fig:potphi}. The value of self-coupling 
constant $\lambda$ is 0.14 at the electroweak scale \cite{Cheong:2021vdb,Mohammedi:2022qqj}. Here the influence of additional parameters 
(such as $\xi$, B, e-folds $N_e$)  on observable parameters  ($n_s,r$) is investigated 
 keeping $\lambda=0.14$ constant \cite{Cheong:2021vdb,Mohammedi:2022qqj}. \\
We plotted this newly defined potential along with the field for different values of $\xi$. Fig.~\ref{fig:potphi} shows 
that  \eqrf{redpot} produces  well-behaved inflationary potentials for arbitrary values of $\xi$. \\
The expressions for the slow roll parameters $\epsilon_1$ and $\epsilon_2$ from \eqrf{ep1}, \eqrf{ep2}, and \eqrf{redpot} are obtained as
\be
\epsilon_1 = \frac{32 M_{Pl}^4}{B M_{Pl}^2  \lambda \phi^6 + B \lambda \xi \phi^8}
\label{ep1f}
\ee
\be
\epsilon_2 = \frac{96M_{Pl}^4+128M_{Pl}^2\xi\phi^2}{BM_{Pl}^2\lambda \phi^6+B \lambda \xi \phi^8}
\label{ep2f}
\ee

\begin{figure}[H]
\centering
\includegraphics[height=7cm,width=7.85cm]{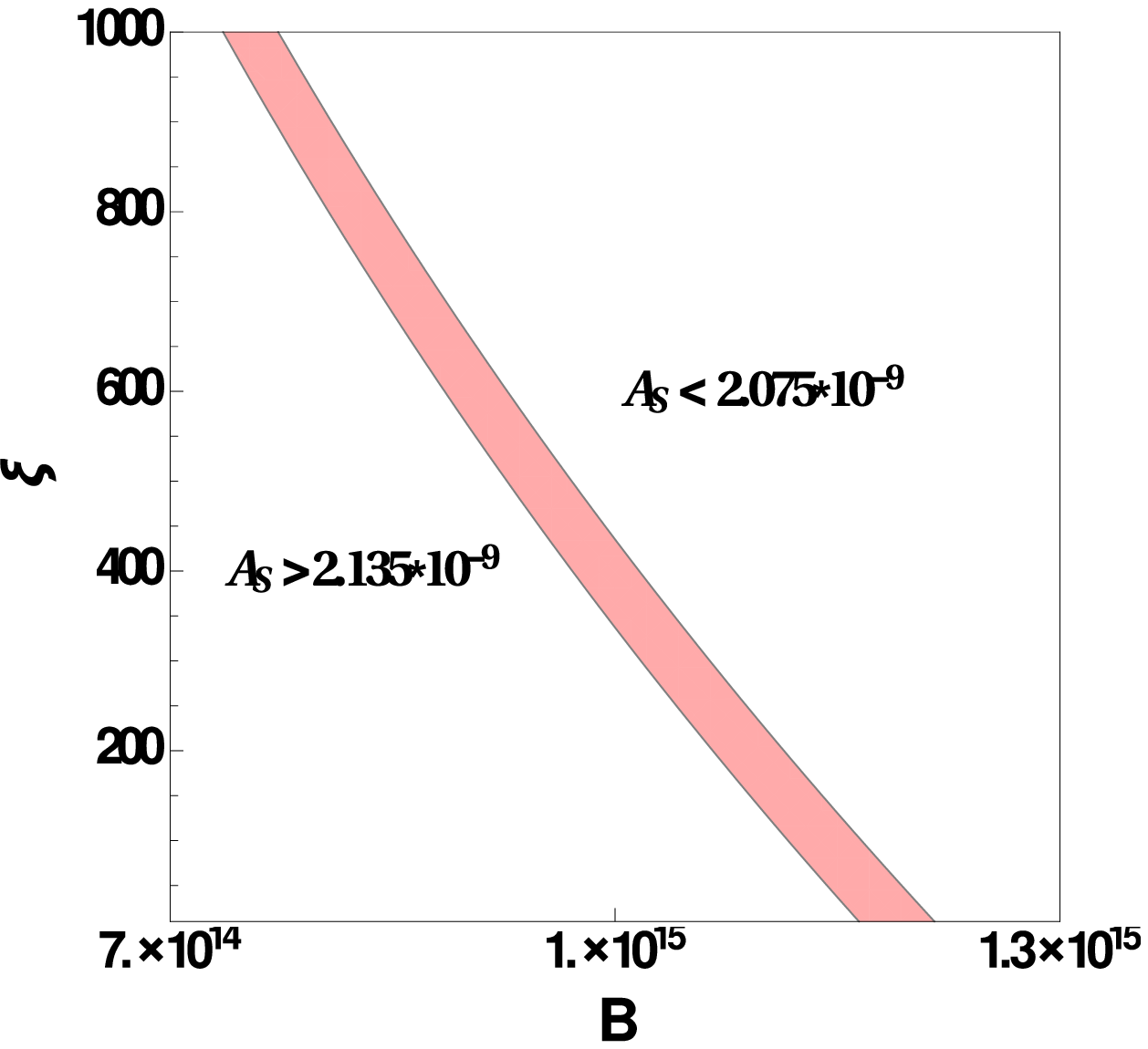}
\hspace{0.5cm}
\includegraphics[height=7cm,width=7.85cm]{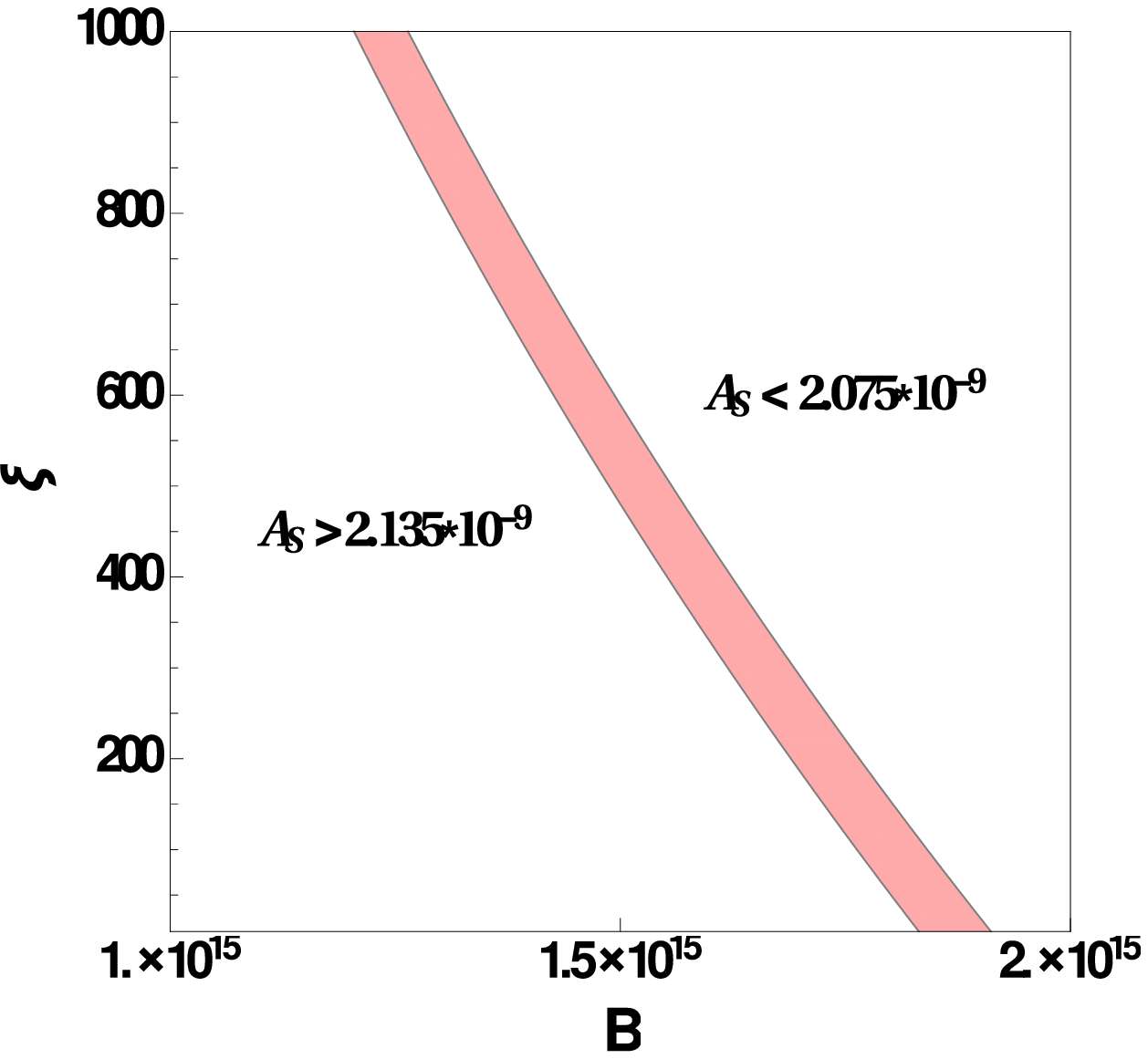}
\caption{These contour plots represent the amplitude of scalar perturbations $A_S$ as function of 
parameters $\xi$ and $B$. the left panel is for $N_k$( number of e-folds)= 50 and right one is for $N_k= 60$
.The  filled  pink region  represents the Planck-2018 constraints on $A_s$.}
\label{fig:prxib}
\end{figure}

Using these slow-roll parameters we compute the amplitude of scalar power spectrum (\ref{scapwr}), scalar spectral index
(\ref{nsv}) and tensor-to-scalar ratio (\ref{rv}) for $N_k=50$ and $60$ for various range of parameters $\xi$ and $B$. As mentioned
earlier we keep $\lambda=0.14$\cite{Cheong:2021vdb,Mohammedi:2022qqj} throughout our calculations.
The variation of scalar amplitude $A_S$ with these parameters is shown in Fig.~\ref{fig:prxib}
The left side of the contour plots in  Fig.~\ref{fig:prxib} is for $N_e=50$ and the right side  is for $N_e=60$. 

We have also shown the allowed range for $A_S$ from  Planck data \cite{Planck:2018vyg}, i.e.,
 $(2.105 \pm 0.030)\times 10^{-9}$ at $68\%$CL. 
 By imposing this constraint, we see that B varies from $7.7\times 10^{14}$ to $1.2\times 10^{15}$ for $N_e=50$,  
and from $1.2\times 10^{15}$ to $1.9\times 10^{15}$ for $N_e=60$,
when $\xi$ decreases from $1000$ to  $10$. We can conclude from Fig.~\ref{fig:prxib} 
that the smaller values of $\xi$ requires larger values of $B$ to produce the same amplitude of scalar perturbations. 
However, $B$ changes by one order of magnitude to satisfy observational constraints on amplitude of scalar perturbations by 
increasing $N_k$ from $50$ to $60$. 
  
The inclusion of the  parameter, $B$,  into  the model  helps to solve the unitarity problem of the original Higgs inflation model, in
which a large value of $\xi$ is required $\left(10^{3-4}\right)$ at $\lambda=0.14$ \cite{Cheong:2021vdb,Mohammedi:2022qqj}. In this model by using a large value of $B$, 
we can minimize the value of $\xi$ to $\mathcal{O}\left(10\right)$.

\begin{figure}[ht]
\centering
\includegraphics[height=7cm,width=7.85cm]{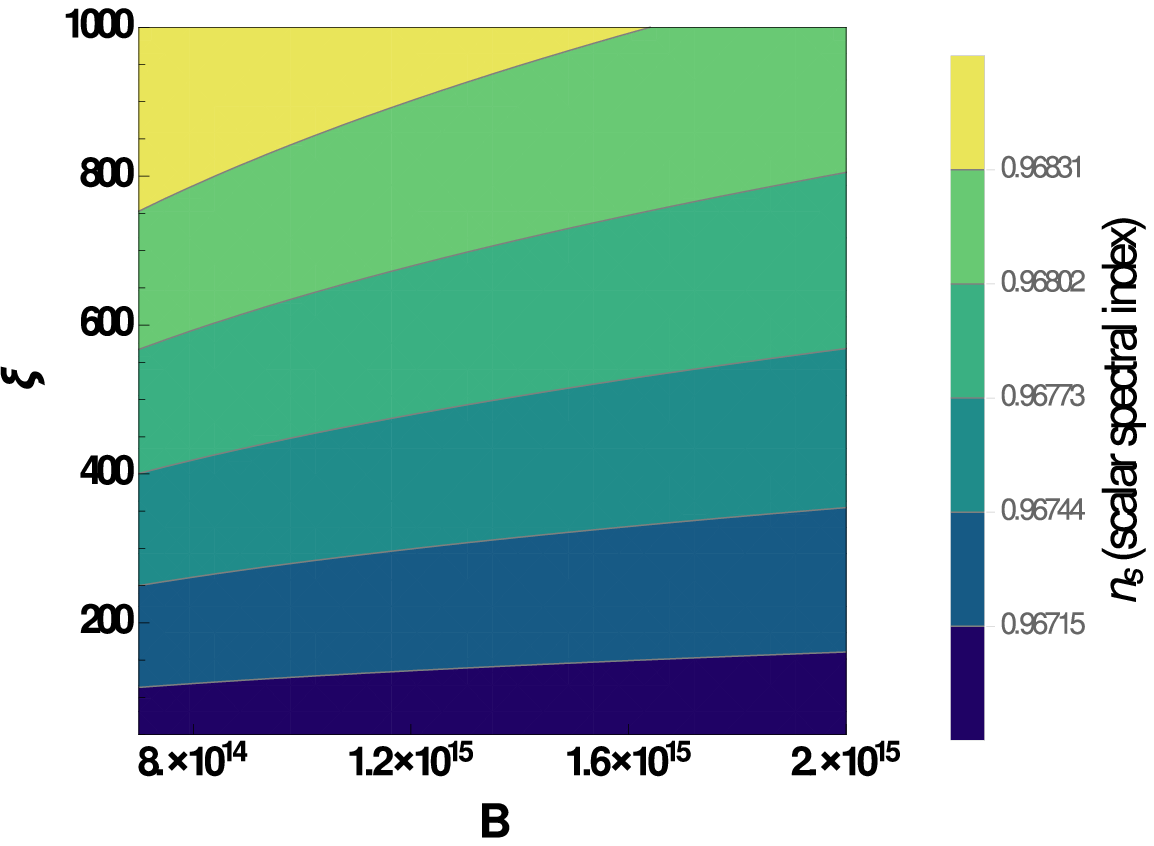}
\hspace{0.5cm}
\includegraphics[height=7cm,width=7.85cm]{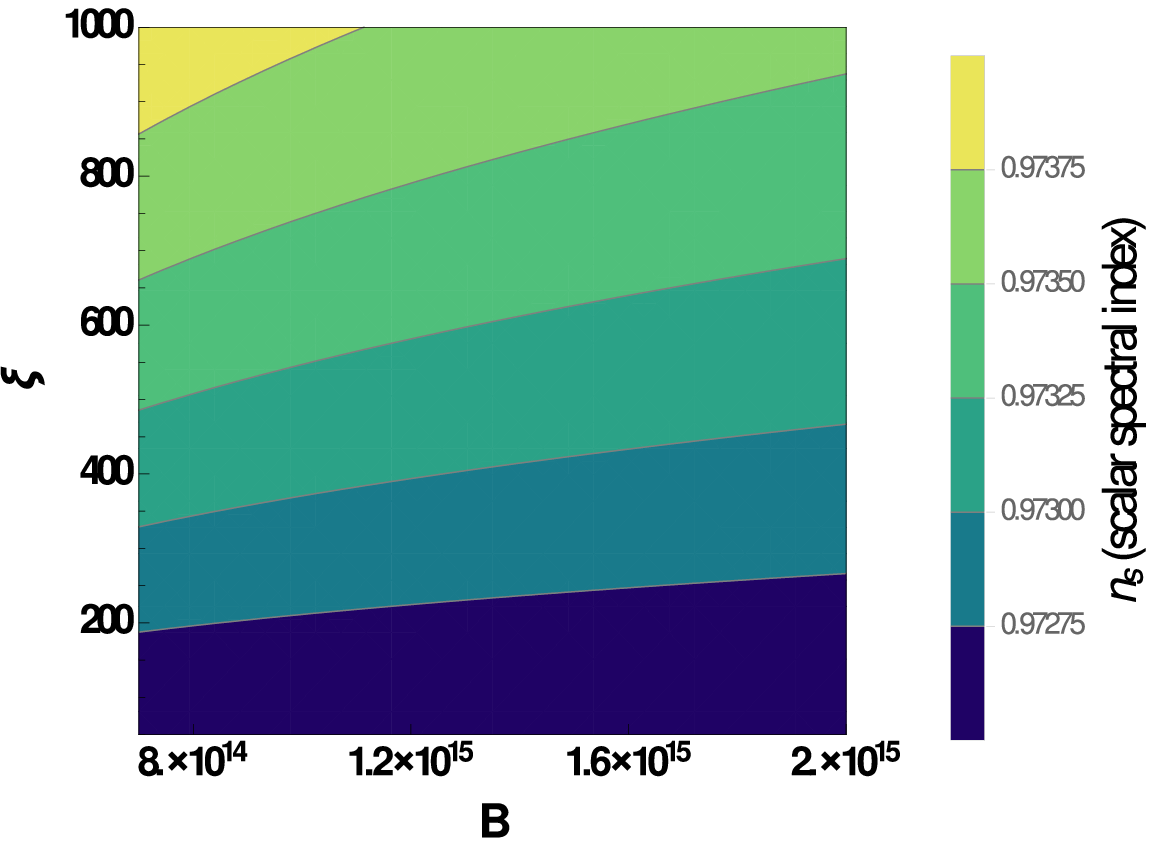}
\caption{The  contour plots represent the scalar spectral index $n_s$ as function of parameters $\xi$ and $B$.
The number of e-foldings $N_k=50$ for the left panel and $N_k=60$ for the right one.}
\label{fig:nsxib}
\end{figure}

\begin{figure}[H]
\centering
\includegraphics[height=7cm,width=7.85cm]{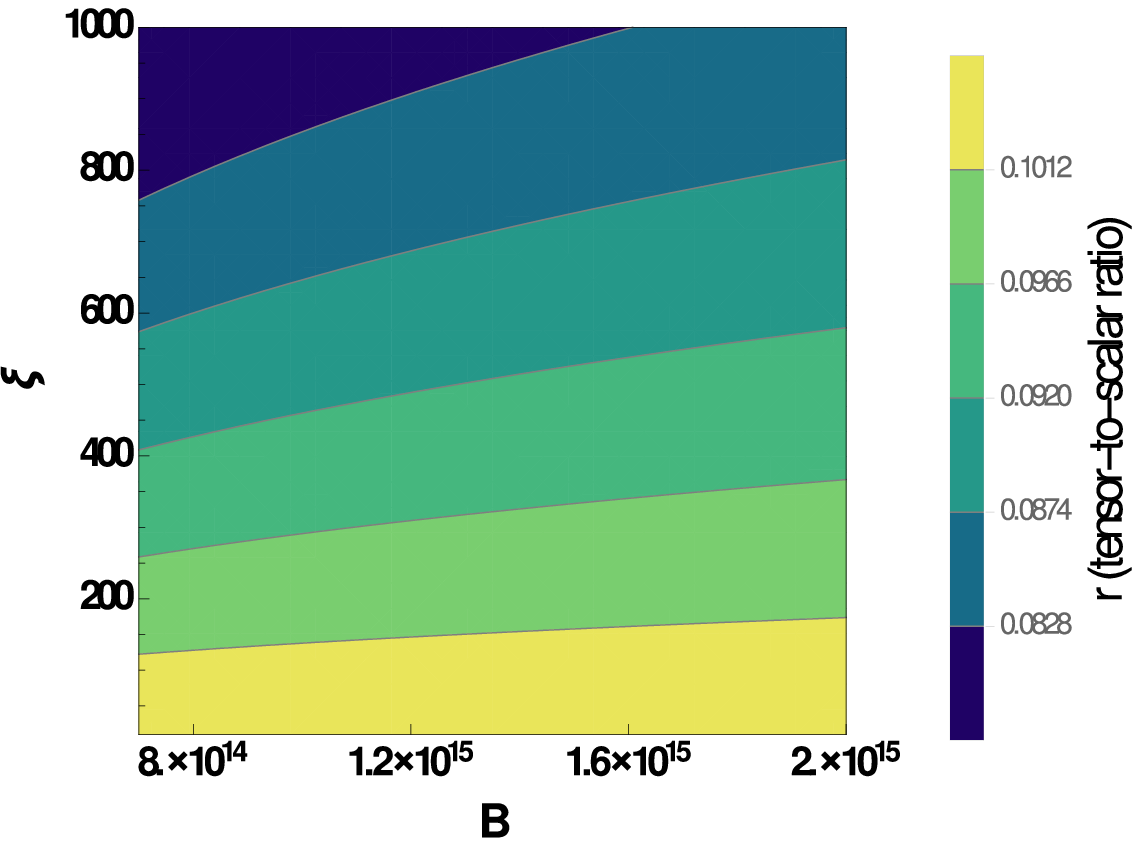}
\hspace{0.5cm}
\includegraphics[height=7cm,width=7.85cm]{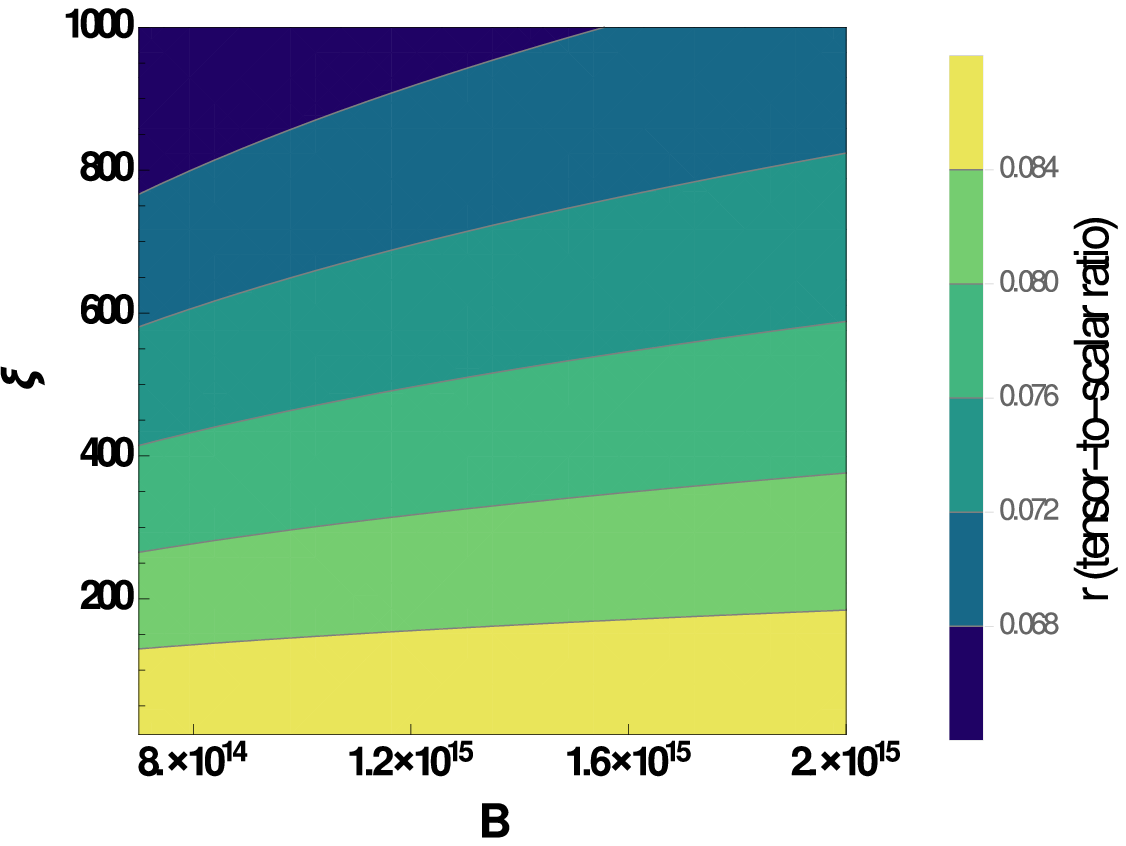}
\caption{The contour plots represents the variation of tensor-to-scalar ratio $r$ as function of 
parameters $\xi$ and $B$. Again, the number of e-foldings $N_k=50$ for the left panel and $N_k=60$ for the right one.}
\label{fig:rxib}
\end{figure}

For the range of $B$ and $\xi$ obtained using constraints  on amplitude of scalar perturbations, the variation of scalar 
spectral index $n_s$ and $r$ is depicted in Fig.~\ref{fig:nsxib} and Fig.~\ref{fig:rxib}.
From these figures it can be seen that the spectral index $n_s$ and tensor-to-scalar ratio $r$ does not change
significantly with $B$ for small values of $\xi$, and this behavior is independent of number of e-foldings. 
It is evident from Fig.~\ref{fig:nsxib} that,
  for constant $B$, $n_s$ increases with $\xi$ and the number of e-foldings. 
It can be seen from Fig.~\ref{fig:rxib} that, for fixed $B$, $r$ decreases with  increase in $\xi$ and also increase in the
number of e-foldings. 

To compute the variation of tensor-to-scalar ratio $r$ with the scalar spectral index $n_s$ for various values of potential
parameters $B$, $\xi$ and the number of e-foldings $N_k$, we solve the background equations 
(\ref{hprime}) and (\ref{eqmn}) numerically 
in terms of the independent variable $N_e=\ln a$. 
The values of the slow-roll parameters $\epsilon_1$ and $\epsilon_2$,  obtained by solving
background equations are then substituted in \eqrftr{rv}{nsv}{cs2} to find $r$ as a function of $n_s$.

\begin{figure}[H]
\centering
\includegraphics[height=7cm,width=15cm]{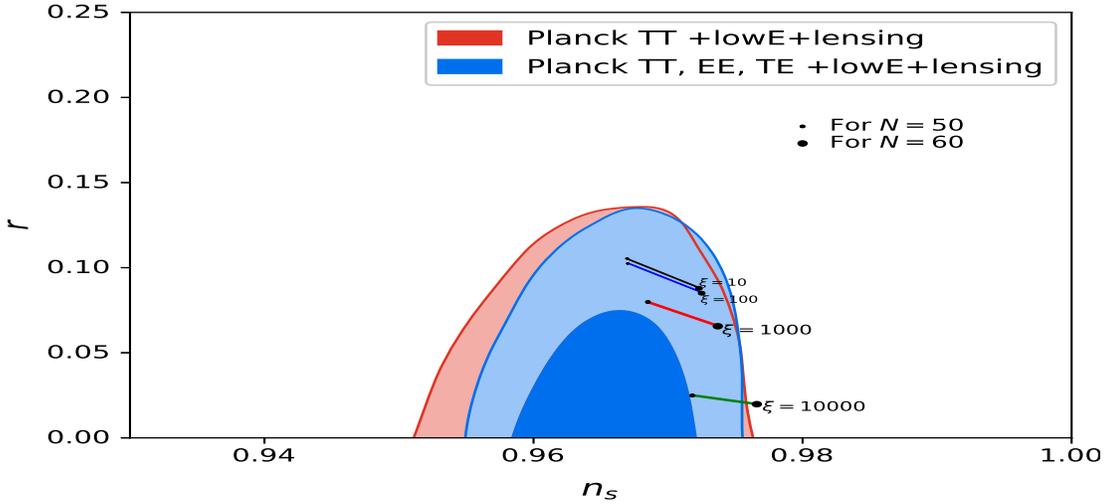}
\caption{r vs $n_s$ predictions for Higgs inflation potential with non-minimally coupled tachyon field along with 
the joint $68\%\,\,C.L.$ and $95\%\,\,C.L.$ Planck-2018 constraints.}
\label{fig:rns1}
\end{figure}
 
The variation of $r$ with respect to $n_s$ along with the Planck-2018 data is 
is shown in Fig.~\ref{fig:rns1}. It is evident from the figure  that for the  non-canonical Higgs inflation model, 
the spectral index $n_s$ and tensor-to-scalar ratio $r$ are within the $95\%$ C.L. of the  Planck 2018 data for smaller
values of $\xi$. 
As the value of $\xi$ increases, the tensor-to-scalar ratio $r$  gets smaller, while the value of 
the spectral index $n_s$ increases. The  values of $n_s$ and $r$ for the original non-minimally 
coupled Higgs inflation model are independent of $\lambda$ and $\xi$, however, in our case, 
both observable parameters depend on the self-coupling constant $\lambda$ and the non-minimal coupling constant $\xi$.

The values of $n_s$ and $r$ for some selected values of $\xi$ and $N_k=50$ and $60$ are also shown in 
Table~\ref{tab:my-table}.
\begin{table}
\centering
\begin{tabular}{|P{2cm}|P{2cm}|P{3cm}|P{3cm}|}
\hline
$\xi$                   & $N_k$ & $n_s$    & r      \\[5pt] \hline
\multirow{2}{*}{10}   & 50   & 0.9669 & 0.1054 \\[5pt] \cline{2-4} 
                      & 60   & 0.9723 & 0.0879 \\[5pt] \hline
\multirow{2}{*}{$10^2$} & 50 & 0.9670 & 0.1024 \\[5pt]\cline{2-4}
                      & 60   & 0.9725 & 0.0852 \\[5pt] \hline
\multirow{2}{*}{$10^3$} & 50 & 0.9686 & 0.0798 \\[5pt] \cline{2-4}
                      & 60   & 0.9738 & 0.0656 \\[5pt] \hline
\multirow{2}{*}{$10^4$} & 50 & 0.9719 & 0.0249 \\[5pt]\cline{2-4} 
                      & 60   & 0.9766 & 0.0199 \\[5pt] \hline
\end{tabular}
\caption{The table shows  the spectral index $n_s$ and tensor-to-scalar ratio $r$ for some selected 
values of $\xi$ and $N_k$ for constant  $B= 10^{15}$.}
\label{tab:my-table}
\end{table}
It can be seen from the table that the value of tensor-to-scalar is large for smaller values of $\xi$. To satisfy 
joint constraints on $r$ from Planck and BICEP3 \cite{BICEP:2021xfz} i.e., $r<0.036$ at $95\%\,\, C. L.$, 
the value of $\xi$ should be of the 
order of $10^4$. Higgs inflation with non-canonical kinetic term yields larger values of tensor-to-scalar $r$ compared to 
the original Higgs inflation model.

\section{Conclusions} \label{conclusion}
Higgs inflation \cite{Bezrukov:2007ep}, where the Higgs field of the standard model is non-minimally coupled to gravity,
is one of the best suited models of inflation from Planck-2018 observations \cite{Planck:2018jri} 
as it predicts smaller tensor-to-scalar ratio. However, this model requires large value of non-minimal coupling constant,
($\xi \sim 10^4$), to satisfy the observational constraints on the amplitude of scalar perturbations, This large value of 
$\xi$ violates unitarity, as the Higgs-Higgs scattering via graviton exchange at a scale $\Lambda= M_{Pl}/\sqrt{\xi}$ becomes 
strongly coupled \cite{Burgess:2009ea,Barbon:2009ya}.

In this work we consider Higgs-inflation with the non-canonical kinetic term of  DBI form along with the
non-minimal coupling to examine whether this can reduce the non-minimal coupling constant. 
As the field $\phi$ has the dimension of mass, a new parameter $B$ with dimension mass\textsuperscript{-4} is introduced 
in the DBI term for dimensional consistency.
To analyze the inflationary dynamics in the Einstein frame we  perform a conformal rescaling of the metric field, which
transforms the Jordan frame action to the minimally coupled action with a non-canonical kinetic term with modified 
Higgs-potential (\ref{redpot}).
   
We find that the Higgs inflation exhibits different features with non-canonical kinetic term. 
The amplitude of the scalar perturbation depends only on the self-coupling constant $\lambda$ and the non-minimal  
coupling constant $\xi$ in the canonical non-minimally coupled Higgs inflation \cite{Bezrukov:2007ep}, 
and to satisfy the cosmological and particle physics requirements at the electroweak scale, 
the order of $\xi$ must be $10^{3-4}$. However, for Higgs-inflation with a non-canonical kinetic term, 
the amplitude of the scalar perturbations $P_s$ is not only a function of $\lambda$ and $\xi$, but also 
depends on the new parameter $B$. At the electroweak scale, the self-coupling constant $\lambda$ is a function of both the 
Higgs mass and the mass of the top quark and has the value $0.14$ \cite{Cheong:2021vdb,Mohammedi:2022qqj}. At this value of $\lambda$, by considering a large value 
of $B$, we can minimize the $\xi$ at ${\mathcal{O}{(10)}}$ to satisfy the constraints on amplitude of scalar perturbations.
 
 The variation of  tensor-to-scalar ratio $r$ and the scalar spectral index $n_s$ 
along with the Joint constraints from Planck 2018 data are shown in Fig.~\ref{fig:rns1} . 
From this, we find that the Higgs inflationary potential with a 
non-canonical kinetic term yields the larger  value of the tensor-to-scalar ratio $r$ and satisfies the Planck-2018 
constraints for $\xi\sim 10^3$ within $95\%\,\, C.L.$.
In case of Higgs-inflation with canonical kinetic term both $n_s$ and  $r$ depend only on  the number of e-foldings
, however, for Higgs-inflation with non-canonical kinetic term, these parameters also depend on 
field-gravity coupling constant $\xi$ and $B$. For selected values of $\xi$ and $N_k$ the values of $n_s$ and $r$ are
shown in Table \ref{tab:my-table}. The bounds on $r$ from from joint analysis of Planck and  
BICEP3 \cite{BICEP:2021xfz} requires the parameter $\xi\sim 10^4$. 
Hence a non-canonical kinetic term of DBI  for Higgs-inflation does not alter the observational constraints on $\xi$ 
significantly.  This implies that the issue of unitarity violation is not resolved even after considering  Higgs inflation
with non-canonical kinetic term. The non-minimally coupled k-inflation with the DBI form of the kinetic energy and 
the potentials derived from string theory has been  considered in \cite{Piao:2002nh, Chingangbam:2004ng,Sen:2009fkx}.
Although quartic potential with $\lambda=0.14$ satisfies the observational constraints for $\xi=10^4$, our analysis
can have phenomenological implications and a detailed analysis can be performed to find the best suited values of $\lambda$ 
and $\xi$ from CMB and LSS observations.

\end{document}